\documentclass[a4paper,12pt]{article}

\usepackage{amsmath}
\usepackage{amssymb}
\usepackage{amsthm}
\usepackage{mathrsfs}
\usepackage[T1]{fontenc}
\usepackage{enumerate}
\usepackage{fullpage}
\usepackage{color}

\def\mtr{matri}
\def\eqn{equation}
\def\cond{condition}

\def\tfn{transformation}

\def\fn{function}
\def\sm{sigma model}

\def\JL{Jacobi--Lie}
\def\jltpy{Jacobi--Lie T-plurality}

\def\dd{Drinfel'd double}

\def\4diml{four-dimensional}
\def\bkg{background}

\def\crspto{corresponding to\ }

\def\-1{^{-1}}
\def\half{\frac{1}{2}}
\def\coor{coordinate}
\def\real{\mathbb{R}}

\def\cd{{\mathfrak d}}
\def\cg{{\mathfrak g}}
\def\tcg{\tilde{\mathfrak g}}

\def\wh{\widehat}

\def\sm{sigma model}
\def\PL{Poisson--Lie }
\def\pltp{Poisson--Lie T-pluralit}

\def\mJL{warped Jacobi--Lie }
\def\JLtp{Jacobi--Lie T-pluralit}

\def\sugra{Supergravity Equation}

\def\cf{{\mathcal {F}}}

\newcommand{\DDp}{DD$^+$}
\newcommand{\Exp}[1]{\operatorname{e}^{#1}}

\newcommand{\abs}[1]{\lvert {#1} \rvert}

\newcommand{\Lie}{\pounds}

\newcommand{\unit}{\mathbf{1}}
\newcommand{\nul}{\mathbf{0}}

\newcommand{\G}{\mathscr{G}}
\newcommand{\tG}{\widetilde{\mathscr{G}}}

\newcommand{\cB}{\mathcal B}

\newcommand{\cE}{\mathcal E}\newcommand{\cF}{\mathcal F}
\newcommand{\cG}{\mathcal G}\newcommand{\cH}{\mathcal H}

\begin{document}
%\title{Alternative spectator versions of Jacobi-Lie T-plurality}
\title{On spectator dependence of Jacobi-Lie T-plurality}

\author{Ivo Petr\footnote{ivo.petr@fit.cvut.cz}
%\\ {\em Department of Applied Mathematics,}
\\ {\em Faculty of Information Technology,}
\\ {\em Czech Technical University in Prague,}
%\\ {\em Th\' akurova 9, 160 00, Prague 6,}
\\ {\em Czech Republic}
\and
Ladislav Hlavat\'y\footnote{ladislav.hlavaty@fjfi.cvut.cz}
%\\ {\em Department of Physics,}
\\ {\em Faculty of Nuclear Sciences and Physical Engineering,}
\\ {\em Czech Technical University in Prague,}
%\\ {\em B\v rehov\' a 7, 115 19, Prague 1,}
\\ {\em Czech Republic}}
\maketitle

\abstract{
Recently introduced Jacobi--Lie T-plurality turned out to be a solution-generating technique in string theory. Being based on Leibniz algebras instead of Drinfeld doubles, it can be understood as a generalization of \pltp y. In this paper we investigate Jacobi--Lie T-plurality with spectators, and focus particularly on modification of the spectator fields by an arbitrary \fn\ $f$. Using this modification new sigma model \bkg s can be constructed as Jacobi--Lie models. For low-dimensional Leibniz algebras  classified a few years ago and warping factor $f(\omega)=\gamma\, e^{\omega}$ we find sigma model backgrounds satisfying \sugra s and check that their plurals again satisfy \sugra s.
}

\tableofcontents

%%%%%%%%%%%%%%%%%%%%%%%%%%%%%%%%%%%%%%%%%%%%%%%%%%%%%%%%%%%%%%%%%%%%%%%%%%%%%%%
%% Paper
%%%%%%%%%%%%%%%%%%%%%%%%%%%%%%%%%%%%%%%%%%%%%%%%%%%%%%%%%%%%%%%%%%%%%%%%%%%%%%%

\section{Introduction}

Nonlinear sigma models satisfying supplementary conditions are used in string theory to model behavior of strings propagating in curved backgrounds %. The \bkg s are 
given by metric $\cG$ and Kalb--Ramond field $\cB$. To define a consistent conformally invariant quantum field theory the background fields in the NS-NS sector have to satisfy one-loop beta function equations \cite{fratsey, calmar} that read
\begin{align}\label{betaG}
0 &= R_{\mu\nu}-\frac{1}{4}H_{\mu\rho\sigma}H_{\nu}^{\ \rho\sigma}+2\nabla_{\mu}\nabla_\nu\Phi,\\ \label{betaB}
0 &=-\frac{1}{2}\nabla^{\rho}H_{\rho\mu\nu}+\nabla^{\rho}\Phi \, H_{\rho\mu\nu},\\
\label{betaPhi} 0 &=
R-\frac{1}{12}H_{\rho\sigma\tau}H^{\rho\sigma\tau}+4\nabla_{\mu}\nabla^\mu\Phi-4\nabla_\mu\Phi \nabla^\mu\Phi .
\end{align}
Here the scalar field $\Phi$ is the dilaton, $\nabla$ denotes covariant derivative, $R_{\mu\nu}$ and $R$ are Ricci tensor and scalar curvature of the metric $\cG$, and $H = \mathrm{d} \cB$ is torsion. Equations \eqref{betaG}--\eqref{betaPhi} are also known as Supergravity (SUGRA) Equations.

Useful tools for finding new solutions of \sugra s are (non-)Abelian T-duality \cite{buscher:ssbfe,recver,aagbl} and \pltp y  \cite{klise,unge:pltp}. Since T-duality/plurality relates backgrounds with different curvature and torsion properties, one may obtain non-trivial \bkg\ from the flat one using plurality transformation. We shall use this approach in one of the examples. In certain cases, however, the plural models do not satisfy equations \eqref{betaG}--\eqref{betaPhi}, and solve generalized  Supergravity Equations \cite{sugra2,Wulff:2016tju,dualnikorejci,hlape:bianchisugra}. In this paper we focus only on solutions of standard SUGRA equations \eqref{betaG}--\eqref{betaPhi}.

In the paper \cite{rezaseph} the authors introduced \jltpy\ as a generalization of \pltp y. Recently it was reformulated in terms of Double Field Theory in Ref. \cite{melsaka}, where Leibniz algebras \DDp\ satisfying further conditions were identified as the algebraic structure suitable for construction of \JL\ models. It was also shown that \jltpy\ transforms solutions of SUGRA equations to solutions of (generalized) SUGRA equations.

Using the classification of low-dimensional \JL\ bialgebras given in \cite{rezaseph:class}, we were able to find solutions of (generalized) \sugra s in three and four dimensions via \jltpy\ \cite{hlape:JLgsugra,pehla:pp as JL}. In these papers we have included spectator fields in the standard way known from \pltp y. It turns out that beside the standard version there is a modification, mentioned e.g. in \cite{geiss}, that can be used for generating solutions of the \sugra s as well. Description of this alternative and its application to solution of \sugra s is the goal of the present paper.

The structure of the paper is the following. In Section \ref{sec:algebra} we give an overview of algebraic structures behind \jltpy. In Section \ref{sec:JLmodels} we summarize the construction of \JL\ models and discuss possible ways to include spectator fields. In Section \ref{examples} we give examples of four-dimensional \JL\ models that satisfy \sugra s and contain either one or two spectators.

\section{Leibniz algebras DD$^+$}\label{sec:algebra}

\JL\ T-plurality is based on isomorphisms of $2D$-dimensional Leibniz algebras DD$^+$ introduced in \cite{melsaka}. Denoting their generators $T_A$ and the product by $\circ$, they are given by structure constants $X_{AB}{}^C$ as
\begin{equation}
T_A \circ T_B = X_{AB}{}^C \, T_C. \label{dd_st_const}
\end{equation}
Beside the Leibniz identities
$$ T_A\circ(T_B\circ T_C)=(T_A\circ T_B)\circ T_C +T_B\circ(T_A\circ T_C)$$ 
one requires that a symmetric bilinear form $\langle ., . \rangle$ can be introduced on DD$^+$, such that there are two Lie subalgebras $\cg$ and $\tcg$ that are maximally isotropic with respect to the form. Denoting generators of $\cg$ and $\tcg$ as $T_a$ and $T^a$ for $a = 1,\ldots,D$ and $T_A=(T_a,\ T^a),$ the form can be chosen as
\begin{equation}\label{eta}
\langle T_A, T_B \rangle = \eta_{AB}, 
\qquad 
\eta_{AB} = \begin{pmatrix} 0 & \delta_a^b \\ \delta_a^b & 0 \end{pmatrix}.
\end{equation}
In terms of $T_a,\ T^a$, the products in the Leibniz algebras DD$^+$ are
\begin{align}
 T_{a}\circ T_{b} &= f_{ab}{}^c\,T_{c}\,, \qquad
 T^a\circ T^b = f_c{}^{ab}\,T^c \,, \label{ddplus1}
\\
 T_{a}\circ T^b &= \bigl(f_a{}^{bc} + 2\,\delta_a^b\,Z^c -2\,\delta_a^c\,Z^b\bigr)\,T_c - f_{ac}{}^b\,T^c +2\,Z_a\,T^b\,, \label{ddplus}
\\
 T^a\circ T_b &= - f_b{}^{ac}\,T_{c} +2\,Z^a\,T_b + \bigl(f_{bc}{}^a +2\,\delta^a_b\,Z_c -2\,\delta^a_c\,Z_b\bigr)\,T^c\,.\label{ddplus3}
\end{align}
Structure constants ${f_{ab}}^c, {f_c}^{ab}, Z_a, Z^a$ satisfy conditions following from Leibniz identities on DD$^+$ and the fact that $\cg$ and $\tcg$ are supposed to be Lie algebras.

There is one-to-one correspondence between Leibniz algebras $DD^+$ and Jacobi-Lie bialgebras introduced in \cite{rezaseph}, and we can use the classification of low-dimensional Jacobi--Lie bialgebras given in \cite{rezaseph:class}.  As explained in Ref. \cite{melsaka}, for construction of \JL\ models depending only on \coor s $x^m$ of the group $\G$ corresponding to algebra $\cg$ we can only use algebras DD$^+$ where $Z^a=0$. In this paper we focus on low-dimensional Type 1 algebras, i.e. those with with $f_b{}^{ba}=Z^a=0$ that lead to sigma models satisfying \sugra s\footnote{Type 2 algebras, i.e. those with $f_b{}^{ba}\neq 0,Z^a=0$ were investigated in Ref. \cite{hlape:JLgsugra}, where they were used to obtain solutions of generalized \sugra s.}. They are displayed in Tables \ref{Table3} and \ref{tab:algebras}.

In the papers \cite{hlape:JLgsugra,pehla:pp as JL} we have found isomorphisms among the low-dimensional algebras $DD^+$, meaning that there exist matrices $C$ that transform algebraic relations \eqref{dd_st_const} generated by $T_A$ to those generated by $\hat T_A = C_A{}^B \, T_B$ with product
\begin{equation}\label{Cplurality}
\hat T_A\circ \hat T_B=\hat X_{AB}{}^C \,\hat T_C.
\end{equation}
Since \eqref{eta} has to hold as well, the conditions on $C$ are
\begin{equation} \label{tfnX}
C_{A}{}^{F}C_{B}{}^{G} X_{FG}{}^H = \hat X_{AB}{}^D\,C_{D}{}^{H}, \qquad C_{A}{}^{F}C_{B}{}^{G} \eta_{FG}=\eta_{AB}.
\end{equation}
We call the algebras given by the coefficients $X_{FG}{}^H$ and $\hat X_{AB}{}^D$ isomorphic or equivalent if there is  a matrix $C$ that solves the \eqn s \eqref{tfnX}. Examples are given in the Appendix.

\section{\JL\ models and alternatives of \jltpy}\label{sec:JLmodels}

Construction of \JL\ models and \JLtp ies that we are going to modify was described in \cite{melsaka} and reviewed in \cite{pehla:pp as JL}. Here we give just a short summary. 

\subsection{Atomic case}

In the absence of spectator fields the \bkg\ fields $\cG,\cB$ are given by constant matrix $E_0$ and structure of Leibniz algebra DD$^+$. We parametrize elements $g$ of the Lie group $\G$  corresponding to Lie algebra $\cg$  as 
$$
g = e^{x^1 T_1} e^{x^2 T_2} \ldots e^{x^D T_D} \in \G.
$$
Matrix $M_A{}^B$ defined via group action on DD$^+$ as
\begin{equation*}
g\-1 \triangleright T_A = M_A{}^B T_B
\end{equation*}
then has the form
\begin{equation}\label{matMAB}
M_A{}^B = 
\begin{pmatrix} 
a_a^b & 0 \\
-\pi^{ac}\,a_c^b & \Exp{-2\Delta} (a\-1)^b_a
\end{pmatrix}
\end{equation}
and defines $\pi^{ac}$ and $\Delta$. The metric $\cG$ and the $\cB$-field of \JL\ model on Lie group $\G$ then can be expressed as symmetric and antisymmetric parts of tensor field 
\begin{align}\label{JLmtz}
 \cF_{mn} = \Exp{-2\omega} \,r_m^a (E_0\-1 +\pi)_{ab}^{-1}\,r_n^b 
\end{align}
where $r_m^a$ are components of right-invariant one-form $\mathrm{d}g g\-1,\ g \in \G$. The factor
$$
\Exp{-2\omega} = \Exp{-2\Delta}\tilde{\sigma}
$$
may in general depend on dual coordinates $\tilde x_m$, but for algebras DD$^+$ of Type 1 and 2, where $Z^a=0$, we can choose $\tilde{\sigma}=1$ and $\omega=\Delta$. Then $\partial_a \omega = \partial_a \Delta = Z_a$.
The condition for pluralizability of \bkg\ $\cF$ given in Ref. \cite{melsaka} reads
\begin{equation}
 \Lie_{v_a} \cF_{mn} + 2\,Z_a\, \cF_{mn} = - \tilde{\sigma}^{-1}\bigl(f_a{}^{bc}+2\,\delta_a^b\,Z^c-2\,\delta_a^c\,Z^b\bigr)\,\cF_{mp}\,v_b^p\,v_c^q\,\cF_{qn}
\label{eq:Lie-cE}
\end{equation}
where $\Lie$ denotes Lie derivative, $v_b^p$ are components of left-invariant vielbeins of the group $\G$, and $f_a{}^{bc}$ are structure constants of the algebra $\tcg$ corresponding to the group $\tG$. For $Z_a = Z^a = 0$ and $\tilde{\sigma}=1$ we may recognize the condition for \PL\ dualizability \cite{klise} of \bkg\ $\cF$. 

Denoting components of left-invariant one-form $g\-1\mathrm{d}g$ as $l_m^a$, the standard dilaton $\Phi$ can be found as
\begin{equation}\label{eq:dilaton-JL0}
\Exp{-2\Phi} =\frac{1}{\sqrt{\abs{\det\cG_{mn}}}}\Exp{-2\varphi} \Exp{-\Delta} \,\abs{\det(l_m^a)}.
\end{equation}
Knowledge of the  \fn\ $\varphi$ is complementary to knowledge of the dilaton.

Contrary to the atomic case, where the equation \eqref{JLmtz} can be used for finding \bkg\ fields, in the presence of spectators we apply DFT formalism \cite{hullzw,hohuzw} with generalized metric $\cH_{MN}$ related to $\cG$ and $\cB$ by 
\begin{equation}\label{HinBG}
 \cH_{MN} \equiv \begin{pmatrix} \cG_{mn} -\cB_{mp}\,\cG^{pq}\,\cB_{qn} & \cB_{mp}\,\cG^{pn} \\ -\cG^{mp}\,\cB_{pn} & \cG^{mn}  \end{pmatrix} .
\end{equation}
The DFT metric has the form
\begin{equation}\label{HMN}
 \cH_{MN}(x) = \cE_M{}^A(x)\, {\cH}_{AB} \,\cE_N{}^B(x)
\end{equation}
where $\cH_{AB}$ is a constant $(2D\times 2D)$-matrix. $\cE_{M}{}^{A}(x)$ is the inverse of 
\begin{align}\label{epsAM}
 \cE_{A}{}^{M}(x)=  \begin{pmatrix} \Exp{\omega} e_a{}^m & 0 \\ -\pi^{ac}\, \Exp{\omega} e_c{}^m & \Exp{-\omega} r^a{}_m \end{pmatrix} 
\end{align}
where $e_m{}^a$ and $r_m{}^a$ are components of the right-invariant vielbein and  one-form on $\G$, i.e. $e_a{}^m r_m{}^b=\delta_a{}^b.$ 
Fields $\cG$, $\cB$ and $\Phi$ constructed above define \JL\ model on the group $\G$. Together with dilaton $\Phi$ we will require them to satisfy the SUGRA equations \eqref{betaG}--\eqref{betaPhi}.

Under plurality \eqref{tfnX} the matrix $\cH_{AB}$ transforms as
\begin{equation}\label{HABhat}
\hat \cH_{AB} = C_A{}^C\, {\cH}_{CD} \, C_B{}^D.
\end{equation}
The background fields $\hat\cG$, $\hat\cB$ of the plural model are then obtained by \eqref{HinBG} from generalized metric $\hat \cH_{MN}$ given by the Eq. \eqref{HMN}.

%calculated from $\hat \cH_{AB}$ and $\hat \cE_{M}{}^{A}$ similarly to \eqref{HinBG} and by 
Plural dilaton $\hat\Phi$ can be found using  transformation of fluxes $F_A$ associated with generalized vielbein $\cE_{A}{}^{M}$ and fuction $\varphi$%DFT dilaton $d$ through 
\begin{align*}%\label{FM}
F_A=%\Exp{-\omega}\cF_A =\Exp{-\omega}\left(\cW^B{}_{AB} + 2\, \cD_A \,d\right)
2\,\Exp{-\omega}\cE_{A}{}^M\,\partial_M \varphi .
\end{align*}
Under \JL\ plurality \eqref{tfnX} the flux transforms as  
\begin{equation}\label{hatFAxdep}
\hat F_A(X)=C_A{}^B F_B(C_\cdot X)
\end{equation}
where $X=(x_1,\ldots,x_D)$.
The plural dilaton $\hat\Phi$ is then found from \eqref{eq:dilaton-JL0}
 and $\hat\varphi$ given by 
\begin{align*}
 \partial_M \hat\varphi = \half \hat \cE_M{}^A \Exp{\hat\omega} \hat F_A.
\end{align*}

\subsection{Standard \JLtp y with spectators}\label{specJLpt}

Extension of \JLtp y to cases with $S$ spectators $y^\alpha$ can be done as follows. We introduce coordinates $$x^M=(x^m,\tilde x_m)=(y^\alpha, x^\mu, \tilde y_\alpha, \tilde x_\mu).$$ The generalized metric \eqref{HMN} can be again written as
\begin{align}\label{HMNspec}
 \cH_{MN} = \cE_M{}^A(x)\, {\cH}_{AB}(y) \,\cE_N{}^B(x),\quad M,N,A,B=1,\ldots,S+D.
\end{align}
The spectator-dependent ${\cH}_{AB}(y)$ is obtained, similarly to \eqref{HinBG}, from symmetric and antisymmetric parts of matrix $E_0(y) = G(y) + B(y)$ as
\begin{equation}
 {\cH}_{AB}(y) = \begin{pmatrix} G_{mn} - B_{mp}\, G^{pq}\, B_{qn} & B_{mp}\, G^{pn} \\ - G^{mp}\, B_{pn} & G^{mn}  \end{pmatrix}.
\end{equation}
The matrix  $\cE_{A}{}^{M}(x)$ now has the form
\begin{align}\label{speceAM}
 \cE_{A}{}^{M}(x)=  \begin{pmatrix} \varepsilon_a{}^m & 0 \\ -\Pi^{ac}\, \varepsilon_c{}^m & \rho^a{}_m \end{pmatrix}
\end{align}
with $(S+D)\times(S+D)$ blocks
\begin{equation}\label{Dging}
\varepsilon_a{}^m =\begin{pmatrix}\unit_S & 0 \\ 0 & \Exp{\omega} e_a{}^m(x)  \end{pmatrix}, \quad \Pi^{ac}(x) =\begin{pmatrix}\nul_S&0 \\ 0&\pi^{ac}(x) \end{pmatrix} , \qquad \varepsilon_a{}^m \rho_m{}^b=\delta_a{}^b.
\end{equation}
The metric $\cG$ and  the field  $\cB$ are again calculated from the relation \eqref{HinBG}, and we get
\begin{equation}\label{standardF}
\cf_{mn}=\cG_{mn}+\cB_{mn}=\rho(x)_{m}{}^{a}(x)\left(E_0(y)\-1+\pi(x)\right)^{-1}_{ab}\rho(x)_{n}{}^{b}(x).
\end{equation}

Formula for \JL\ plurality for the spectator dependent matrix $E_0(y)$  reads
\begin{equation}\label{E0spec}
\wh E_0(y)=\left(\left(\mathcal{P}+ E_0(y) \cdot \mathcal{R}\right)^{-1} \cdot \left(\mathcal{Q}+E_0(y) \cdot \mathcal{S}\right)\right)^T,
\end{equation}
where the matrices $\mathcal P,\mathcal Q,\mathcal R,\mathcal S $ are obtained by extension of the $D\times D$ blocks $P,Q,R,S$ of the \tfn\ \mtr x $C^{-1}$
$$C^{-1}=\begin{pmatrix}
 P & Q \\
 R & S
\end{pmatrix}, \qquad T_A = (C^{-1})_A{}^B \, \hat T_B $$
to $(S+D)\times (S+D)$ matrices
\begin{equation*}%\label{pqrs2}
\mathcal{P} =\begin{pmatrix}\unit_S &0 \\ 0&P \end{pmatrix}, \quad \mathcal{Q} =\begin{pmatrix}\nul_n&0 \\ 0&Q \end{pmatrix}, \quad \mathcal{R} =\begin{pmatrix}\nul_S&0 \\ 0&R \end{pmatrix}, \quad \mathcal{S} =\begin{pmatrix}\unit_S &0 \\ 0& S \end{pmatrix}.
\end{equation*}

In \cite{melsaka,saka2} dilaton was suggested in the form
\begin{equation}\label{specdil393}
 \Exp{-2\Phi(x,y)} = \frac{1}{\sqrt{\abs{\det \cG(x,y)}}}\Exp{-2\hat d(y)}\Exp{-2\varphi(x)} \Exp{-\Delta(x)}  \abs{\det \left( l_m{}^a(x)\right)}.
%\label{eq:specdilaton-JL1}
\end{equation}
To get solutions of \sugra s we have to choose 
\begin{equation}\label{dilmelsaka}
\hat d(y)=-\frac{1}{4}\ln\frac{(\det E_0(y))^2}{\det E_{0S}(y)}, \quad E_{0S}(y)=\half(E_0(y)+E_0^T(y)).
\end{equation}
Examples of standard \JLtp y in dimension 1+3 were given in \cite  {pehla:pp as JL}.

\subsection{Modification of \JLtp y with spectators}

One can check that the standard \JL\ \bkg\  \eqref{standardF} does not satisfy the pluralizability condition \eqref{eq:Lie-cE}, and we can try to find its modification whose special case is the condition for the \bkg\  \eqref{standardF}. The modification consists in replacing the spectator part in \eqref{Dging} by the so-called \emph{warping factor} $f(\omega(x))$, see \cite{geiss,sibylla},
\begin{equation}\label{modif genDging}
\varepsilon(x)=\begin{pmatrix}f(\omega(x))\, \unit_S & 0 \\ 0 & \Exp{\omega(x)} e(x)  \end{pmatrix}
\end{equation}
where $f$ is arbitrary differentiable \fn. Let us write the modified \bkg\ as 
$$
\cf_{mn}(x)=J(\omega)_{m}{}^{p}F_{pq}(x) J(\omega)_{n}{}^{q},
$$
or
\begin{equation}\label{standardF2}\cf(x)=J(x)\cdot F(x)\cdot J(x)^T,
\end{equation}
where
$$ J(\omega)=\begin{pmatrix}\frac{1}{f(\omega)}\ \unit_S & 0 \\ 0 & \Exp{-\omega}\unit_D \end{pmatrix}.$$
Then 
\begin{align*}
 \Lie_{v_a} \cF(x)  = 
\partial_a \omega
\left(\frac{dJ(\omega)}{d\omega}\cdot  F(x)\cdot  J^T+J\cdot F(x)\cdot \frac{dJ^T(\omega)}{d\omega}\right)+ J\cdot \Lie_{v_a} F (x)\cdot J^T,
\end{align*}
where
$$ DJ(\omega)=\frac{dJ(\omega)}{d\omega}\cdot J(\omega)\-1 =\begin{pmatrix}-\frac{f'(\omega)}{f(\omega)}\ \unit_S & 0 \\ 0 & -\unit_D \end{pmatrix}.$$

Matrix $F(x)$ is of the form \eqref{standardF}, where 
\begin{equation}\label{dging1}
\rho_m{}^a =\begin{pmatrix}\unit_S & 0 \\ 0 & r_m{}^a(x)  \end{pmatrix}, \quad \Pi^{ac}(x) =\begin{pmatrix}\nul_S&0 \\ 0&\pi^{ac}(x) \end{pmatrix},
\end{equation}
and satisfies Poisson-Lie pluralizability condition \cite{klim:proc}
\begin{equation}\label{kseqnPL}
 \Lie_{v_a} F_{mn}(x)  = -F_{mp} v_b{}^p f_a{}^{bc}v_c{}^q F_{qn},\quad  m,n,p,q=1,\ldots, S+D,\quad  a,b,c=1,\ldots,D,
\end{equation} 
or
\begin{equation}\label{kseqn}
 \Lie_{v_a} F  = -F\cdot v^T\cdot f_a\cdot v\cdot F.
\end{equation} 
Since $\partial_a \omega = Z_a$, we can write
\begin{equation}
 \Lie_{v_a} \cF(x) 
=  Z_a 
\left(DJ\cdot \cf(x)+ \cf(x)\cdot DJ^T \right) - J\cdot \ F\cdot v^T\cdot f_a\cdot  v\cdot F^T \cdot J^T. \\ 
\end{equation}
In addition to that, the components of the left-invariant vielbein are of the form
$$ v_c{}^q=(\nul_{S},v_c{}^\kappa),\quad \kappa=1,\ldots,D,$$
so that $v\cdot J=v$ and
\begin{align*}
 \Lie_{v_a} \cF(x) 
= & Z_a 
\left(DJ\cdot \cf(x)+ \cf(x)\cdot DJ^T \right) - J\cdot \ F\cdot J^T\cdot v^T\cdot f_a\cdot  v\cdot J\cdot F^T  \cdot J^T \\
= & Z_a 
\left(DJ\cdot \cf(x)+ \cf(x)\cdot DJ^T \right) - \cf\cdot v^T \cdot f_a\cdot  v \cdot\cf.
\end{align*}
 
Therefore, \emph{pluralizability condition} for the \JL\ \bkg s of the Type 1 and the Type 2, i.e.  for $\tilde \sigma =1$, reads
\begin{align}\Lie_{v_a}  & \cF_{mn}(x)= \nonumber \\ & Z_a \left(DJ(\omega)_{m}{}^{p}\cF(x)_{pn}+\cF(x)_{mp}DJ(\omega)_{n}{}^{p}\right)-\cF_{mp}\,v_b^p\,f_a{}^{bc}\,v_c^q\,\cF_{qn}\,.
\label{modif eq:Lie-cE}
\end{align}
This condition holds for arbitrary function $f(\omega)$. However, we were able to find only two functions for which the \JL\ models satisfy \sugra s. Either $f(\omega)=1$, i.e the standard \JLtp y  introduced in Sec. \ref{specJLpt}, or  $f(\omega)=\gamma\, e^{\omega}$.

\subsection{Warped \JLtp y}\label{sec:warped}

Setting the warping function as $$f(\omega)=\gamma\, e^{\omega}$$ where $\gamma$ is an arbitrary constant different from zero we get an alternative  to standard \JLtp y  through the extension
\begin{equation}\label{genDging}
\varepsilon(x)=\begin{pmatrix}\gamma\, \Exp{\omega}\ \unit_S & 0 \\ 0 & \Exp{\omega} e(x)  \end{pmatrix}, \quad \Pi(x) =\begin{pmatrix}\nul_S&0 \\ 0&\pi(x) \end{pmatrix}.
\end{equation}
Note that this is not a special case of the standard \JLtp y where  $f(\omega)=1$. Nevertheless, both alternatives are based on the same \JL\ algebra DD$^+$. 

Moreover, we need formula for
dilaton. Using the modified form \eqref{genDging} of the inverse vielbein and 
$$ (A\-1)^T+A\-1=A\-1(A +A^T)A\-1 $$  we find that 
$$ \det \cG(x,y)= 1/\gamma^{2n}
 \Exp{-2(D+n)\,\Delta(x)}\det r(x)^2 
\det\left(E_0(y)\-1+\pi(x)\right)^{-2}\frac{\det E_{0S}(y)}{\det E0(y)^2}. $$ From the Eq. \eqref{specdil393} we then  get 
formula for spectator-dependent dilaton (cf. (3.7) in \cite{melsaka})
\begin{align}\label{specdil37}
\Exp{-2\Phi(x,y)} =& \gamma^n
 \Exp{-2\varphi(x)+(D+n)\,\Delta(x)}\det a(x) 
\det\left(E_0(y)\-1+\pi(x)\right)\frac{\det E0(y)}{\sqrt{\det E_{0S}(y)}}.
\end{align}

\section{Examples of warped \JLtp y}\label{examples}
\subsection{Four-dimensional flat models corresponding to the algebra $\{1|1\}$ and one spectator}\label{11spec}

Let us  start with flat torsionless model obtained as \mJL\ model given by the algebra $\{1|1\}$ and some $E_0(t)$\footnote{We have $ n=1$ and $y_1=t$.}. As follows from the Table \ref{tab:algebras}, the algebras $\cg, \tcg$ are Abelian with ${f_{ab}}^c={f_c}^{ab}=0$ and the only non-zero component of $Z_a$ is $Z_1=\frac{1}{2}$. For this algebra $\omega=\frac{x_1}{2}$ and $\pi^{ab} = 0$. After a bit tedious calculations we can find flat \JL\ \bkg\
\begin{equation}\label{mtz11}
 \cf(t,x)=  \left(
\begin{array}{cccc}
 -H'(t)^2 e^{-x_1} & 0 & 0 & 0 \\
 0 & e^{-x_1} & 0 & 0 \\
 0 & 0 & e^{H(t)-x_1} & 0 \\
 0 & 0 & 0 & e^{H(t)-x_1} \\
\end{array}
\right)
\end{equation}
generated by
\begin{equation}\label{E011}
E_0(t)= \left(\begin{array}{cccc}
-\gamma^2 H'(t)^2 & 0 & 0 & 0 \\
 0 & 1 & 0 & 0 \\
 0 & 0 & e^{H(t)} & 0 \\
 0 & 0 & 0 & e^{H(t)} \\
\end{array}
\right)
\end{equation} 
and arbitrary function $H(t)$ with $H'(t)\neq 0$. Usual Supergravity Equations for this \bkg\ are satisfied with vanishing dilaton and the formula \eqref{specdil37} yields 
$$\varphi=\frac{3}{4}\,x_1-\half \left( H(t)+\ln{H'(t)}\right).$$

On the other hand, \JL\ \bkg,  obtained by standard construction presented in Sec. \ref{specJLpt} with the same $E_0(t)$, reads
$$ \cf_S(t,x)= \left(
\begin{array}{cccc}
-\gamma^2 H'(t)^2 & 0 & 0 & 0 \\
 0 & e^{-x_1} & 0 & 0 \\
 0 & 0 & e^{H(t)-x_1} & 0 \\
 0 & 0 & 0 & e^{H(t)-x_1} \\
\end{array}
\right).$$
It is not flat as its scalar curvature is $R=\frac{1}{2} \left(e^{x_1}-3\right)$.

\subsubsection{Warped \JL\ models corresponding to $\{3;-2|1\}$}

In the following subsections we shall find \JL\ plurals to sigma model given by \bkg\  \eqref{mtz11}, and verify that they satisfy \sugra s.
Structure constants of algebra $\{3;-2|1\}$ can be found in Table \ref{tab:algebras}. Since $Z_1 = -1$, $\omega(x)=-x_1$.
Formula \eqref{E0spec} for the plurality matrix \eqref{11to321}
gives
$$ \hat E_0(t)=\left(
\begin{array}{cccc}
 -\gamma^2 H'(t)^2 & 0 & 0 & 0 \\
 0 & 4 & 0 & 0 \\
 0 & 0 & e^{-H(t)}+\frac{e^{H(t)}}{4} &
   e^{-H(t)}-\frac{e^{H(t)}}{4} \\
 0 & 0 & e^{-H(t)}-\frac{e^{H(t)}}{4} &
   e^{-H(t)}+\frac{e^{H(t)}}{4} \\
\end{array}
\right) $$ 
and from the algebra $\{3;-2|1\}$ we get curved background
$$ \hat \cf(t, x) = \left(
\begin{array}{cccc}
 -e^{2 x_1} H'(t)^2 & 0 & 0 & 0 \\
 0 & 4 e^{2 x_1} & 0 & 0 \\
 0 & 0 & e^{-H(t)-2 x_1}+\frac{1}{4} e^{H(t)+2 x_1} & e^{-H(t)-2 x_1}-\frac{1}{4} e^{H(t)+2 x_1}
   \\
 0 & 0 & e^{-H(t)-2 x_1}-\frac{1}{4} e^{H(t)+2 x_1} & e^{-H(t)-2 x_1}+\frac{1}{4} e^{H(t)+2 x_1}
   \\
\end{array}
\right)$$
with vanishing scalar curvature and torsion.
Together with dilaton
$$ \Phi=-\frac{1}{2}\left( {H(t)}+2 x_1\right)$$
obtained from the formula \eqref{specdil37}, they satisfy Supergravity Equations. 

\subsubsection{Warped \JL\ models corresponding to $\{5;-1|1\}$   }

Formula \eqref{E0spec} for the matrix \eqref{11to511}
gives
$$ \hat E_0(t)=\left(
\begin{array}{cccc}
 -\gamma^2H'(t)^2 & 0 & 0 & 0 \\
 0 & 1 & 0 & 0 \\
 0 & 0 & e^{-H(t)} & 0 \\
 0 & 0 & 0 & e^{-H(t)} \\
\end{array}
\right). $$ 
From the algebra $\{5;-1|1\}$ we get \bkg
$$ \hat \cf(t, x) =\left(
\begin{array}{cccc}
 -e^{x_1} H'(t)^2 & 0 & 0 & 0 \\
 0 & e^{x_1} & 0 & 0 \\
 0 & 0 & e^{-H(t)-x_1} & 0 \\
 0 & 0 & 0 & e^{-H(t)-x_1} \\
\end{array}
\right) $$
that together with dilaton
$$ \Phi=-H(t)-x_1$$
satisfy usual Supergravity Equations. Scalar curvature of the metric is zero and the metric can be brought into the Brinkman form of a plane-parallel wave 
\begin{equation}\label{ppwave51}
ds^2 =\frac{2 \left(z_3^2+z_4^2\right)}{u^2} du^2 + 2dudv + dz_3^2 + dz_4^2  
\end{equation}
using \tfn\ of coordinates
\begin{align*}
t&=H^{(-1)}\left(-\ln \left(\frac{1}{12} \left(-\frac{3
   z_3^2}{u^2}-\frac{3 z_4^2}{u^2}+\frac{6
   v}{u}\right)\right)\right),\\ x_1&=\ln \left(\frac{1}{12}
   \left(-\frac{3 z_3^2}{u^2}-\frac{3
   z_4^2}{u^2}+\frac{6 v}{u}\right)\right)+2
   \ln (u),\\ x_2&=-u z_4, \qquad x_3=-u z_3.
\end{align*}
   
\subsubsection{Warped \JL\ models corresponding to $\{1|2\}$   }

Formula \eqref{E0spec} for the matrix \eqref{11to12} gives
$$ \hat E_0(t)=\left(
\begin{array}{cccc}
 -\gamma^2H'(t)^2 & 0 & 0 & 0 \\
 0 & 1 & 0 & 0 \\
 0 & 0 & \frac{e^{H(t)}}{e^{2 H(t)}+1} &
   \frac{e^{2 H(t)}}{e^{2 H(t)}+1} \\
 0 & 0 & -\frac{e^{2 H(t)}}{e^{2 H(t)}+1} &
   \frac{e^{H(t)}}{e^{2 H(t)}+1} \\
\end{array}
\right). $$ 
From the algebra $\{1|2\}$ we get
$$ \hat \cf(t, x) =\left(
\begin{array}{cccc}
 -e^{-x_1} H'(t)^2 & 0 & 0 & 0 \\
 0 & e^{-x_1} & 0 & 0 \\
 0 & 0 & \frac{e^{H(t)+x_1}}{e^{2 H(t)}+e^{2
   x_1}} & \frac{e^{2 H(t)}}{e^{2 H(t)}+e^{2
   x_1}} \\
 0 & 0 & -\frac{e^{2 H(t)}}{e^{2 H(t)}+e^{2
   x_1}} & \frac{e^{H(t)+x_1}}{e^{2 H(t)}+e^{2
   x_1}} \\
\end{array}
\right) $$
that together with dilaton
$$ \Phi = -\frac{1}{2} \ln \left(e^{2 H(t)}+e^{2 x_1}\right)+x_1$$
and nontrivial torsion satisfy usual Supergravity Equations. 

After coordinate transformation
\begin{align*}
t=&H^{(-1)}\left(-\ln \left(\frac{2 u^5 v-u^4
   \left(z_3^2+z_4^2\right)+2 u
   v+z_3^2+z_4^2}{4 u^2
   \left(u^4+1\right)}\right)\right),\\ x_1=&\ln
   \left(\frac{2 u^5 v-u^4
   \left(z_3^2+z_4^2\right)+2 u
   v+z_3^2+z_4^2}{4 u^2
   \left(u^4+1\right)}\right)+2 \ln
   (u),\\ x_2=&\sqrt{u^2+\frac{1}{u^2}}
   z_4,\quad x_3=\sqrt{u^2+\frac{1}{u^2}}
   z_3
\end{align*}
the pp-wave metric has the form
\begin{equation}\label{ppwave12spec}
ds^2= \frac{2 u^2 \left(u^4-5\right) \left(z_3^2+z_4^2\right)}{\left(u^4+1\right)^2} du^2 + 2dudv + dz_3^2 + dz_4^2,
\end{equation}
and torsion is
$$ H=-\frac{4\,u}{1+u^4}(du\wedge dz_3\wedge dz_4)$$   

\subsubsection{Warped \JL\ models corresponding to $\{5;-1|2\}$}

Formula \eqref{E0spec} for the matrix \eqref{11to512} gives
$$ \hat E_0(t)=\left(
\begin{array}{cccc}
 -\gamma^2H'(t)^2 & 0 & 0 & 0 \\
 0 & 1 & 0 & 0 \\
 0 & 0 & \frac{e^{H(t)}}{e^{2 H(t)}+1} &
   \frac{1}{e^{2 H(t)}+1} \\
 0 & 0 & -\frac{1}{e^{2 H(t)}+1} &
   \frac{e^{H(t)}}{e^{2 H(t)}+1} \\
\end{array}
\right). $$ 
From the algebra $\{5;-1|2\}$ we get
$$ \hat \cf(t, x) =\left(
\begin{array}{cccc}
 -e^{x_1} H'(t)^2 & 0 & 0 & 0 \\
 0 & e^{x_1} & 0 & 0 \\
 0 & 0 & \frac{e^{H(t)+x_1}}{e^{2
   \left(H(t)+x_1\right)}+1} & \frac{1}{e^{2
   \left(H(t)+x_1\right)}+1} \\
 0 & 0 & -\frac{1}{e^{2
   \left(H(t)+x_1\right)}+1} &
   \frac{e^{H(t)+x_1}}{e^{2
   \left(H(t)+x_1\right)}+1} \\
\end{array}
\right) $$
that together with  dilaton
$$ \Phi=-\frac{1}{2} \ln \left(e^{2
   \left(H(t)+x_1\right)}+1\right)$$
and nontrivial torsion satisfy usual Supergravity Equations. In Brinkmann coordinates the metric has the pp-wave form \eqref{ppwave12spec}. 

%{To bring the metric into the Brinkman form it is necessary that  H(t) is invertible.  Then 
%\begin{equation}\label{ppwave52}
%ds^2 =\frac{2 u^2 \left(u^4-5\right)
%   \left(z_3^2+z_4^2\right)}{\left(u^4+1\right)^2}du^2 + 2dudv + dz_3^2 + dz_4^2 ,
%\end{equation}
%%Choosing H(t)=t we get \tfn
%and torsion is
%$$ H=\frac{4\,u}{1+u^4}(du\wedge dz_3\wedge dz_4)$$   
%Corresponding transformation is 
%\begin{align*}
%t=&H^{(-1)}\left(-\ln \left(\frac{2 u^5 v-u^4
%   \left(z_3^2+z_4^2\right)+2 u v+z_3^2+z_4^2}{4 u^2
%   \left(u^4+1\right)}\right)\right),\\ x_1=&\ln \left(\frac{2
%   u^5 v-u^4 \left(z_3^2+z_4^2\right)+2 u v+z_3^2+z_4^2}{4
%   u^2 \left(u^4+1\right)}\right)+2 \ln
%   (u),\\ x_2=&\sqrt{u^2+\frac{1}{u^2}}
%   z_4,\quad x_3=\sqrt{u^2+\frac{1}{u^2}} z_3.
%\end{align*}

\subsection{Four-dimensional models corresponding to the algebra $\{2|1\}$ and one spectator}\label{21spec}

Algebra $\{2|1\}$ is given by 
$$
T_2 \circ T_3=T_1
$$
and $Z_3 = \frac{1}{2}$. \JL\ model satisfying \sugra s with vanishing dilaton is obtained from  $\omega=\frac{x_3}{2}$ and 
\begin{equation}
E_0(t) =\left(
\begin{array}{cccc}
 0 & 0 & \gamma\,  h(t) & \gamma\,  (h(t)+1)
   \\
 0 & 0 & 2 & 2 \\
 \gamma\,  h(t) & 2 & 0 & 0 \\
 \gamma\,  (h(t)+1) & 2 & 0 & -t \\
\end{array}
\right),
\end{equation}
where $h(t)$ is arbitrary function. Background of warped \JL\ model then is
\begin{equation}\label{mtz21}
\cf(t,x)=\left(
\begin{array}{cccc}
 0 & 0 & e^{-x_3} h(t) & e^{-x_3} (h(t)+1) \\
 0 & 0 & 2 e^{-x_3} & 2 e^{-x_3} \\
 e^{-x_3} h(t) & 2 e^{-x_3} & 0 & 2 e^{-x_3}
   x_2 \\
 e^{-x_3} (h(t)+1) & 2 e^{-x_3} & 2 e^{-x_3}
   x_2 & -e^{-x_3} \left(t-4 x_2\right) \\
\end{array}
\right).
\end{equation}
The metric is Ricci flat but not flat. Vanishing dilaton gives $\varphi=\frac{3}{4}\,x_3.$

The \JL\ \bkg\ obtained from the same $E_0(t)$ by the standard construction presented in Sec. \ref{specJLpt} reads
$$ \cf_S(t,x)=\left(
\begin{array}{cccc}
 0 & 0 & e^{-\frac{x(3)}{2}} h(t) &
   e^{-\frac{x(3)}{2}} (h(t)+1) \\
 0 & 0 & 2 e^{-x(3)} & 2 e^{-x(3)} \\
 e^{-\frac{x(3)}{2}} h(t) & 2 e^{-x(3)} & 0 & 2
   e^{-x(3)} x(2) \\
 e^{-\frac{x(3)}{2}} (h(t)+1) & 2 e^{-x(3)} & 2
   e^{-x(3)} x(2) & -e^{-x(3)} (t-4 x(2)) \\
\end{array}
\right).$$
It is not Ricci flat and satisfies \sugra s  for $h(t)=0$ and the dilaton
$$ \Phi(x)=-e^{-\sqrt{e^{-x_3}}} \text{Ei}\left(\sqrt{e^{-x_3}}\right)-\frac{x_3 }{2}$$
with the exponential integral function $\text{Ei}$ defined as
$$\text{Ei}(x)=\int_{-\infty}^x \frac{e^t}{t} d t.$$
   
In the following subsections we shall find \JL\ plurals to sigma model given by \bkg\  \eqref{mtz21}.Plural model \crspto plurality $\{2|1\}\cong \{3;-2|2\}$ does not exist because matrix $\hat E_0$ given by \eqref{E0spec}  does not exist for the isomorphisms given by \eqref{21to322}.

\subsubsection{Warped \JL\ models corresponding to $\{4;-1|1\}$}

Formula \eqref{E0spec} for the plural matrix \eqref{21to411}
gives
$$ \hat E_0(t)=\left(
\begin{array}{cccc}
 0 & -\gamma  & \frac{1}{2} \gamma  h(t)
   & 0 \\
 -\gamma  & -t & 0 & 1 \\
 -\frac{1}{2} \gamma  h(t) & 0 & 0 &
   -\frac{1}{2} \\
 0 & -1 & -\frac{1}{2} & 0 \\
\end{array}
\right). $$ 
From the algebra $\{4;-1|1\}$ we get $\omega=\frac{x_1}{2}$ and
$$ \hat \cf(t, x) =
\left(
\begin{array}{cccc}
 0 & -e^{x_1} & \frac{h(t)}{2} & 0 \\
 -e^{x_1} & -t \,e^{x_1} &
   x_1 & 1 \\
 -\frac{h(t)}{2} & -x_1 & -e^{-x_1} x_1
   & -\frac{e^{-x_1}}{2} \\
 0 & -1 & -\frac{e^{-x_1}}{2} & 0 \\
\end{array}
\right) $$
that together with dilaton
$\Phi=-x_1$
satisfy usual Supergravity Equations. Torsion and scalar curvature vanish.

\subsubsection{Warped \JL\ models corresponding to $\{2.i|2\}$}

Formula \eqref{E0spec} for the matrix \eqref{21to2i1} gives
$$ \hat E_0(t)=\left(
\begin{array}{cccc}
 0 & \gamma -\gamma  h(t) & 0 & \gamma  h(t) \\
 \frac{1}{3} \gamma  (h(t)+3) & -t & \frac{2}{3} & 0 \\
 0 & -2 & 0 & 2 \\
 -\frac{1}{3} \gamma  h(t) & 0 & -\frac{2}{3} & 0 \\
\end{array}
\right). $$ 
From the algebra $\{2.i|2\}$ we get \bkg
$$ \hat \cf(t, x) =\left(
\begin{array}{cccc}
 0 & -\frac{h(t)-2 e^{-x_1}+1}{2-e^{x_1}} & 0 &
   \frac{h(t)}{2-e^{x_1}} \\
 \frac{h(t)+2 e^{-x_1}+1}{e^{x_1}+2} & t
   \left(-e^{-x_1}\right) & \frac{2}{e^{x_1}+2} & \frac{2
   x_1}{e^{x_1}+2} \\
 0 & \frac{2}{e^{x_1}-2} & 0 & -\frac{2}{e^{x_1}-2} \\
 -\frac{h(t)}{e^{x_1}+2} & \frac{2 x_1}{e^{x_1}-2} &
   -\frac{2}{e^{x_1}+2} & -\frac{4 e^{x_1} x_1}{e^{2 x_1}-4}
   \\
\end{array}
\right) $$
with nontrivial  torsion. Together with dilaton
$$ \Phi=-\frac{1}{2} \ln \left(e^{2 x_1}-4\right)+x_1$$
they satisfy usual Supergravity Equations. Scalar curvature vanishes and the only non-trivial component of Ricci tensor is $R_{22}=\frac{4 \left(5 e^{2 x_1}+4\right)}{\left(e^{2 x_1}-4\right){}^2}$.

\subsubsection{Warped \JL\ models corresponding to $\{4;-1|2\}$   }

Formula \eqref{E0spec} for the matrix \eqref{21to2i1} gives
$$ \hat E_0(t)=\left(
\begin{array}{cccc}
 0 & -\frac{1}{3} \gamma  (h(t)+3) & \frac{1}{3} \gamma  h(t)
   & 0 \\
 \gamma  (h(t)-1) & -t & 0 & 2 \\
 -\gamma  h(t) & 0 & 0 & -1 \\
 0 & -\frac{2}{3} & -\frac{1}{3} & 0 \\
\end{array}
\right). $$ 
From the algebra $\{4;-1|2\}$ we get \bkg
$$ \hat \cf(t, x) =\left(
\begin{array}{cccc}
 0 & -\frac{e^{x_1} \left(h(t)+2 e^{x_1}+1\right)}{2
   e^{x_1}+1} & \frac{e^{x_1} h(t)}{2 e^{x_1}+1} & 0 \\
 \frac{e^{x_1} \left(h(t)-2 e^{x_1}+1\right)}{2 e^{x_1}-1} & t
   \left(-e^{x_1}\right) & \frac{2 e^{x_1} x_1}{2 e^{x_1}-1} &
   \frac{2 e^{x_1}}{2 e^{x_1}-1} \\
 \frac{e^{x_1} h(t)}{1-2 e^{x_1}} & -\frac{2 e^{x_1} x_1}{2
   e^{x_1}+1} & \frac{4 e^{x_1} x_1}{1-4 e^{2 x_1}} &
   \frac{1}{1-2 e^{x_1}} \\
 0 & -\frac{2 e^{x_1}}{2 e^{x_1}+1} & -\frac{1}{2 e^{x_1}+1} &
   0 \\
\end{array}
\right) $$
that together with dilaton
$$ \Phi=-\frac{1}{2} \ln \left(1-4\,e^{2 x_1}\right)$$
satisfy usual Supergravity Equations. Scalar curvature vanishes and the only non-trivial component of Ricci tensor is $R_{22}=\frac{4 e^{2 x_1} \left(4 e^{2 x_1}+5\right)}{\left(1-4 e^{2 x_1}\right){}^2}$.

\subsubsection{Warped \JL\ models corresponding to $\{4.iii;1|2\}$   }

Formula \eqref{E0spec} for the matrix \eqref{21to2i1} gives
$$ \hat E_0(t)=\left(
\begin{array}{cccc}
 0 & \gamma -\gamma  h(t) & \gamma  h(t) & 0 \\
 \frac{1}{3} \gamma  (h(t)+3) & -t & 0 & -\frac{2}{3} \\
 -\frac{1}{3} \gamma  h(t) & 0 & 0 & -\frac{1}{3} \\
 0 & 2 & -1 & 0 \\
\end{array}
\right). $$ 
From the algebra $\{4.iii;1|2\}$ we get \bkg\ with torsion
$$ \hat \cf(t, x) =\left(
\begin{array}{cccc}
 0 & -\frac{h(t)-2 e^{-x_1}+1}{2-e^{x_1}} &
   \frac{h(t)}{2-e^{x_1}} & 0 \\
 \frac{h(t)+2 e^{-x_1}+1}{e^{x_1}+2} & t
   \left(-e^{-x_1}\right) & \frac{2 x_1}{e^{x_1}+2} &
   -\frac{2}{e^{x_1}+2} \\
 -\frac{h(t)}{e^{x_1}+2} & \frac{2 x_1}{e^{x_1}-2} & -\frac{4
   e^{x_1} x_1}{e^{2 x_1}-4} & -\frac{e^{x_1}}{e^{x_1}+2} \\
 0 & -\frac{2}{e^{x_1}-2} & \frac{e^{x_1}}{e^{x_1}-2} & 0 \\
\end{array}
\right) $$
that together with dilaton
$$ \Phi=-\frac{1}{2} \ln \left(e^{2 x_1}-4\right)+x_1$$
satisfy usual Supergravity Equations. Scalar curvature vanishes and the only non-trivial component of Ricci tensor is $R_{22}=\frac{4 \left(5 e^{2 x_1}+4\right)}{\left(e^{2 x_1}-4\right){}^2}$.

\subsection{Four-dimensional models corresponding to algebra $[1|1]$ and two spectators}

To verify formulas in Section \ref{sec:warped} it is desirable to check them also for more than one spectator. In the following we show an example of \jltpy\ with two spectators $y_1,y_2$ and model  \crspto two-dimensional \JL\ bialgebra.

\JL\ bialgebra $[1|1]$ is formed by Abelian Lie algebras $\cg, \tcg$ and cocycle $\phi_0 = 2 Z_a T^a = T^{1}+T^{2}$, see the Table \ref{Table3}. Choosing
\begin{equation}
E_0(y_1,y_2) =\left(
\begin{array}{cccc}
 1 & 0 & -y_2 & 0 \\
 0 & 1 & y_1 & 0 \\
 -y_2 & y_1 & y_1^2+y_2^2-1 & 0 \\
 0 & 0 & 0 & 1 \\
\end{array}
\right), \quad \varphi=\frac{1}{2}(x_1+x_2)
\end{equation}
we get sigma model \bkg\ and dilaton
\begin{equation}
\cf(x,y)= \Exp{-x_1-x_2}\left(
\begin{array}{cccc}
 \frac{1}{\gamma ^2} & 0 & -\frac{y_2}{\gamma } & 0 \\
 0 & \frac{1}{\gamma ^2} & \frac{y_1}{\gamma } & 0 \\
 -\frac{y_2}{\gamma } & \frac{y_1}{\gamma } & y_1^2+y_2^2-1 & 0 \\
 0 & 0 & 0 & 1 \\
\end{array}
\right), \quad \Phi=-\frac{1}{4}(x_1+x_2)
\end{equation}
satisfying Supergravity Equations. The metric has vanishing scalar curvature and the non-trivial components of Ricci tensor read $R_{33}=R_{34}=R_{43}=R_{44}=-\frac{1}{2}$.
%Kretschmann scalar, but we were not able to bring it to Brinkmann form.
%$$ Ricci=\{\{0, 0, 0, 0\}, \{0, 0, 0, 0\}, \{0, 0, -1/2, -1/2\}, \{0, 0, -1/2, -1/2\}\}$$.

\subsubsection{\JL\ model corresponding to $[2;-1|1]$}

Isomorphism $ [1|1]\cong [2;-1|1]$  given by \eqref{c1121}  
transforms $E_0(y_1,y_2)$ to 
$$ \hat E_0(y_1,y_2)=\frac{1}{y_1^2+y_2^2}\left(
\begin{array}{cccc}
 y_1^2 & y_1 y_2 & y_2 & y_2 \\
 y_1 y_2 & y_2^2 & -y_1 & -y_1 \\
 -y_2 & y_1 & 1 & -y_1^2-y_2^2+1 \\
 y_2 & -y_1 & y_1^2+y_2^2-1 & y_1^2+y_2^2-1 \\
\end{array}
\right) $$
and corresponding \jltpy\  gives \bkg 
\begin{equation}\label{cf2spectators}
\hat \cf(x,y)=\frac{\Exp{x_2}}{y_1^2+y_2^2}\left(
\begin{array}{cccc}
 \frac{y_1^2}{\gamma ^2} & \frac{y_1 y_2}{\gamma ^2} &
   \frac{y_2}{\gamma } & \frac{\left(x_1+1\right) y_2}{\gamma
   } \\
 \frac{y_1 y_2}{\gamma ^2} & \frac{y_2^2}{\gamma ^2} &
   -\frac{y_1}{\gamma } & -\frac{\left(x_1+1\right)
   y_1}{\gamma } \\
 -\frac{y_2}{\gamma } & \frac{y_1}{\gamma } & 1 &
   x_1-y_1^2-y_2^2+1 \\
 -\frac{\left(x_1-1\right) y_2}{\gamma } &
   \frac{\left(x_1-1\right) y_1}{\gamma } & x_1+y_1^2+y_2^2-1
   & x_1^2+y_1^2+y_2^2-1 \\
\end{array}
\right)
\end{equation}
with nontrivial torsion and scalar curvature. Nevertheless, this \bkg\ together with dilaton
$$\Phi(x,y)=-\frac{1}{4}\,x_2-\half \ln \left(y_1^2+y_2^2\right),$$ obtained from \eqref{specdil37}, satisfy \sugra s.

\section{Conclusions}

In this paper we have defined a modification of the \jltpy\ with spectators replacing the inverse vielbein $\varepsilon$ in  the standard construction \eqref{HMNspec}--\eqref{Dging} by its $f$-dependent form \eqref{modif genDging}. We have presented the modified pluralizability condition \eqref{modif eq:Lie-cE} for arbitrary differentiable \fn\ $f(\omega)$. Let us note that the modified condition  with $f(\omega)=\gamma\, e^{\omega}$ is equal to the unmodified condition \eqref{eq:Lie-cE} for the atomic case. This is not true for the standard \JLtp y. 

We have also shown that the warped \jltpy\ is a solution-generating technique to the Supergravity equations similarly to the standard one. Focusing on Type 1 algebras, i.e. those given by \eqref{ddplus1}--\eqref{ddplus3}  with ${f_b}^{ba} = Z^a = 0$, we have used isomorphisms among four- and six-dimensional Type 1 algebras to get solutions of the \sugra s. The isomorphisms allowed us to distinguish Type 1 algebras into several equivalence classes, where \jltpy\ transformation can be applied. For each equivalence class we have found a simple initial \JL\ model that satisfies \sugra s  \eqref{betaG}--\eqref{betaPhi} and used warped \JL\ T-plurality to transform it to more complex \JL\ models  satisfying \sugra s as well. For the warped \JL\ plurality we have found slightly modified formula for dilaton \eqref{specdil37}, which is necessary  to get solutions of the  \sugra s.

Results of the procedures are given in Section \ref{examples} where we present several initial \JL\ models with one and two spectators and simple dilatons as well as their plural counterparts.  Beside these examples the formulas for the warped \jltpy\ were checked for many other isomorphisms of the algebras of both Types 1 and 2. Warped \JL\ models for Type 2 algebras can be found in the Sec. 6.1 of the Ref. \cite{hlape:JLgsugra}.

Due to the particular choice of initial \JL\ models, namely with vanishing or linear dilaton, nearly all of the presented plural models have vanishing scalar curvature and simple Ricci tensor. For some of them we were able to bring them to the Brinkmann form to prove that they are plane-parallel waves. 

Open question is whether only \fn s $f(\omega)=const.$ and $f(\omega)=const.\, e^{\omega} $ are suitable factors for the modified  \jltpy. In spite of intensive effort we were not able to find any other function that could replace them. 

\section{Appendix - Isomorphisms of low-dimensional algebras}\label{sec:iso}

\subsection{Type 1 four-dimensional \JL\ algebras}

Four-dimensional \JL\ algebras were classified in \cite{rezaseph:class}. Their duals needed for construction of \JL\ models are given in the Table \ref{Table3}.

%%%%%%%%%%%%%%%%%%%%%%%%%%%%%%%%%%%%%%%%%%%%%%%%%%%%%%%%%%%%%%%%%%%%%%%%%%%%%%%

\begin{table}
\begin{center}  
\begin{tabular}{c c l l l lp{0.15mm} }
\multicolumn{5}{l}{}\\
\hline
\hline
{\footnotesize Name of $\cd$} &{\footnotesize ${\tcg}$ }& {\footnotesize ${\cg}$}
&{\footnotesize Product definitions of ${\cg}$}&{\footnotesize $\phi_{0} = 2\, Z_a T^a$}&{\footnotesize Comments} \\
\hline
\vspace{2mm}
{\footnotesize $[1\,|\,1]$}&{\footnotesize $I$}&{\footnotesize $I$}&{\footnotesize $T_i \circ T_j=0$}&{\footnotesize $T^{1}+T^{2}$}&\\
\vspace{2mm}{\footnotesize $[2;\alpha\,|\,1]$}&
{\footnotesize $I$}&{\footnotesize $II$}&{\footnotesize $T_1 \circ T_2 = T_1$}&{\footnotesize $\alpha\, T^{2}$}&{\footnotesize $\alpha\in\real\smallsetminus\{0\}$}\\
\hline
\end{tabular}
\caption{Duals of two-dimensional Jacobi-Lie bialgebras in Table 5 of \cite{rezaseph:class}.
\label{Table3}}
\end{center} 
\end{table}
%%%%%%%%%%%%%%%%%%%%%%%%%%%%%%%%%%%%%%%%%%%%%%%%%%%%%%%%%%%%%%%%%%%%%%%%%%%%%%%%%%%%%%%%%%

There are two isomorphisms among these algebras, namely
\begin{align}
[1|1]\cong& [2;-1|1],\\
[2;\alpha|1]\cong &[2;-\frac{\alpha}{\alpha+1}|1].
\end{align}
Corresponding transformation matrices are
\begin{equation}
\label{c1121}C= \left(
\begin{array}{cccc}
 0 & 0 & 0 & 1 \\
 -1 & 0 & 0 & 0 \\
 -1 & 1 & 0 & 0 \\
 0 & 0 & -1 & -1 \\
\end{array}
\right)
\end{equation} 
and
\begin{equation}
\label{c2121}C= \left(
\begin{array}{cccc}
 0 & 0 & 1 & 0 \\
 0 & -\frac{1}{\alpha +1} & 0 & 0 \\
 1 & 0 & 0 & 0 \\
 0 & 0 & 0 & -\alpha -1 \\
\end{array}
\right).
\end{equation}

\subsection{Type 1 six-dimensional \JL\ algebras}

Six-dimensional \JL\ algebras were classified in \cite{rezaseph:class} as well. As mentioned in the Introduction we were interested in models generated by \JL\ algebras with ${f_b}^{ba} = Z^a = 0$. They are displayed in the Table \ref{tab:algebras}. These algebras are duals of those given in \cite{rezaseph:class} with ${f_{ab}}^b=Z_a=0$.

%%%%%%%%%%%%%%%%%%%%%%%%%%%%%%%%%%%%%%%%%%%%%%%%%%%%%%%%%%%%%%%%%%%%%%%%%%%%%%%
\begin{table}
\begin{center}  
\begin{tabular}{c c l l l lp{0.15mm} }
\multicolumn{5}{l}{}\\
\hline
\hline
{\footnotesize Name of $\cd$} &{\footnotesize ${\tcg}$ }& {\footnotesize ${\cg}$}
&{\footnotesize Product definitions of ${\cg}$}&{\footnotesize $\phi_{0} = 2\, Z_a T^a$}&{\footnotesize Comments} \\
\hline
\vspace{2mm}
{\footnotesize $\{1\,|\,1\}$}&{\footnotesize $I$}&{\footnotesize $I$}&{\footnotesize $T_i \circ T_j =0$}&{\footnotesize $T^{1}$}&\\
\vspace{2mm}{\footnotesize $\{2\,|\,1\}$}&
{\footnotesize $I$}&{\footnotesize $II$}&{\footnotesize $T_2 \circ T_3=T_1$}&{\footnotesize $T^{3}$}&\\
\vspace{2mm}{\footnotesize $\{3\,|\,1\}$}&
{\footnotesize $I$}&{\footnotesize {$III$}}&{\footnotesize {$T_1 \circ T_2=-(T_2+T_3),\ T_1 \circ T_3=-(T_2+T_3)$}}&{\footnotesize $T^{3}-T^{2}$}&\\
\vspace{2mm}{\footnotesize $\{3;b\,|\,1\}$}&
{\footnotesize $I$}&{\footnotesize $III$}&{\footnotesize $T_1 \circ T_2=-(T_2+T_3),\ T_1 \circ T_3=-(T_2+T_3)$}& {\footnotesize $b\,T^{1}$}&{\footnotesize $b\in {\real\smallsetminus\{0\}}$}\\
\vspace{2mm}{\footnotesize $\{4;b\,|\,1\}$}&
{\footnotesize $I$}&{\footnotesize $IV$}&{\footnotesize $T_1 \circ T_2=-(T_2-T_3),\ T_1 \circ T_3=-T_3$}& {\footnotesize $b\,T^{1}$}&{\footnotesize $b\in \real\smallsetminus\{0\}$}\\
\vspace{2mm}{\footnotesize $\{5;b\,|\,1\}$}&
{\footnotesize $I$}&{\footnotesize $V$}&{\footnotesize $T_1 \circ T_2=-T_2,\ T_1 \circ T_3=-T_3$}& {\footnotesize $b\,T^{1}$}&{\footnotesize $b\in \real\smallsetminus\{0\}$}\\
\vspace{2mm}{\footnotesize $\{6_0;b\,|\,1\}$}&
{\footnotesize $I$}&{\footnotesize $VI_{0}$}&{\footnotesize $T_1 \circ T_3=T_2,\ T_2 \circ T_3=T_1$}& {\footnotesize $b\,T^{3}$}&{\footnotesize $b>0$}\\
{\footnotesize $\{6_a;b\,|\,1\}$}&{\footnotesize $I$}&{\footnotesize $VI_{a}$}&{\footnotesize $T_1 \circ T_2=-(aT_2+T_3),\ T_1 \circ T_3=-(T_2+aT_3)$}& {\footnotesize $b\,T^{1}$}&{\footnotesize $a>0,a\neq1$}\\
&&&&&{\footnotesize $b\in \real\smallsetminus\{0\}$}\\
\vspace{2mm}{\footnotesize $\{7_0;b\,|\,1\}$}&
{\footnotesize $I$}&{\footnotesize $VII_{0}$}&{\footnotesize $T_1 \circ T_3=-T_2,\ T_2 \circ T_3=T_1$}& {\footnotesize $b\,T^{3}$}&{\footnotesize $b>0$}\\
{\footnotesize $\{7_a;b\,|\,1\}$}&{\footnotesize $I$}&{\footnotesize $VII_{a}$}&{\footnotesize $T_1 \circ T_2=-(aT_2-T_3),\ T_1 \circ T_3=-(T_2+aT_3)$}& {\footnotesize $b\,T^{1}$}&{\footnotesize $a>0$}\\
&&&&&{\footnotesize $b\in \real\smallsetminus\{0\}$}\\
\vspace{2mm}{\footnotesize $\{1\,|\,2\}$}&
{\footnotesize $II$}&{\footnotesize $I$}&{\footnotesize $T_i \circ T_j=0$}&{\footnotesize $T^{1}$}&\\
\vspace{2mm}{\footnotesize $\{2.i\,|\,2\}$}&
{\footnotesize $II$}&{\footnotesize $II.i$}&{\footnotesize $T_1 \circ T_3=T_2$}&{\footnotesize $T^{1}$}&\\
\vspace{2mm}{\footnotesize $\{2.ii\,|\,2\}$}&
{\footnotesize $II$}&{\footnotesize $II.ii$}&{\footnotesize $T_1 \circ T_3=-T_2$}&{\footnotesize $T^{1}$}&\\
\vspace{2mm}{\footnotesize $\{3;b\,|\,2\}$}&
{\footnotesize $II$}&{\footnotesize $III$}&{\footnotesize $T_1 \circ T_2=-(T_2+T_3),\ T_1 \circ T_3=-(T_2+T_3)$}&{\footnotesize $b\,T^{1}$}&{\footnotesize $b\in \real\smallsetminus\{0\}$}\\
\vspace{2mm}{\footnotesize $\{4;b\,|\,2\}$}&
{\footnotesize $II$}&{\footnotesize $IV$}&{\footnotesize $T_1 \circ T_2=-(T_2-T_3),\ T_1 \circ T_3=-T_3$}&{\footnotesize $b\,T^{1}$}&{\footnotesize $b\in \real\smallsetminus\{0\}$}\\
\vspace{2mm}{\footnotesize $\{4.iii;b\,|\,2\}$}&
{\footnotesize $II$}&{\footnotesize $IV.iii$}&{\footnotesize $T_1 \circ T_2=T_2-T_3,\ T_1 \circ T_3=T_3$}&{\footnotesize $b\,T^{1}$}&{\footnotesize $b\in \real\smallsetminus\{0\}$}\\
\vspace{2mm}{\footnotesize $\{5;b\,|\,2\}$}&
{\footnotesize $II$}&{\footnotesize $V$}&{\footnotesize $T_1 \circ T_2=-T_2,\ T_1 \circ T_3=-T_3$}&{\footnotesize $b\,T^{1}$}&{\footnotesize $b\in \real\smallsetminus\{0\}$}\\
\vspace{2mm}{\footnotesize $\{6_0.iii;b\,|\,2\}$}&
{\footnotesize $II$}&{\footnotesize $VI_{0}.iii$}&{\footnotesize $T_1 \circ T_2=T_3,\ T_1 \circ T_3=T_2$}&{\footnotesize $b\,T^{1}$}&{\footnotesize $b >0$}\\
{\footnotesize $\{6_a;b\,|\,2\}$}&{\footnotesize $II$}&{\footnotesize $VI_{a}$}&{\footnotesize $T_1 \circ T_2=-(aT_2+T_3),\ T_1 \circ T_3=-(T_2+aT_3)$}&{\footnotesize $b\, T^1$}& {\footnotesize $a>0,a\neq1$}\\
&&&&&{\footnotesize $b \in \real\smallsetminus\{0\}$}\\
\vspace{2mm}{\footnotesize $\{7_0.i;b\,|\,2\}$}&
{\footnotesize $II$}&{\footnotesize $VII_{0}.i$}&{\footnotesize $T_1 \circ T_2=-T_3,\ T_1 \circ T_3=T_2$}& {\footnotesize $b\,T^{1}$}&{\footnotesize $b>0$}\\
\vspace{2mm}{\footnotesize $\{7_0.ii;b\,|\,2\}$}& {\footnotesize $II$}&{\footnotesize $VII_{0}.ii$}&{\footnotesize $T_1 \circ T_2=T_3,\ T_1 \circ T_3=-T_2$}& {\footnotesize $b\,T^{1}$}&{\footnotesize $b>0$}\\
{\footnotesize $\{7_a;b\,|\,2\}$}&{\footnotesize $II$}&{\footnotesize $VII_{a}$}&{\footnotesize $T_1 \circ T_2=-(aT_2-T_3),\ T_1 \circ T_3=-(T_2+aT_3)$}&{\footnotesize $b\,T^1$}& {\footnotesize$a > 0$}\\
&&&&&{\footnotesize $b \in \real\smallsetminus\{0\}$}\\
\hline
\end{tabular}
\caption{Type 1 three-dimensional Jacobi-Lie bialgebras.They are duals of bialgebras in \cite{rezaseph:class}, Table 7 with $f_b{}^{ba}=0$. $X_{0}=0$.\label{tab:algebras}}
\end{center}
\end{table}
%%%%%%%%%%%%%%%%%%%%%%%%%%%%%%%%%%%%%%%%%%%%%%%%%%%%%%%%%%%%%%%%%%%%%%%%%%%%%%%

We were able to identify following classes of equivalence of Type 1 algebras in  \cite{pehla:pp as JL}:
\begin{equation}\label{1|1}
\{1\,|\,1\}
\cong\{3;-2\,|\,1\} \cong\{5;-1\,|\,1\}
 \cong\{1\,|\,2\} \cong\{5;-1\,|\,2\}.\end{equation}
\begin{align}\label{2|1}
\{2\,|\,1\}
&\cong\{4;-1\,|\,1\}\cong\{2.i\,|\,2\}\cong\{2.ii\,|\,2\} \cong\{3;-2\,|\,2\}\cong\\ &\cong\{4;-1\,|\,2\}\cong\{4iii;1\,|\,2\}.\nonumber
\end{align}
\begin{align}\label{31|1}
\{3;-1|1\}\cong\{3;2|1\}\cong\{6_0;1|1\}\cong\{6_0.iii;1|2\} ,
\end{align}
\begin{align}\label{3b|1}
 b\neq -2,-1,2 : \  \{3;b|1\}&\cong\{3;\frac{-2\,b}{2+b}|1\}\cong\{3;b|2\}\cong\{3;\frac{-2\,b}{2+b}|2\}\cong
\\
\cong &\{6_{b+1};-b|1\}\cong\{6_{\frac{2-b}{2+b}};\frac{2b}{2+b}|1\} \nonumber
\end{align}

\begin{align}\label{4b|1} b\neq -2,-1 :  \{4;b|1\}\cong\{4;\frac{-b}{1+b}|1\}\cong\{4;b|2\}\cong\{4;\frac{-b}{1+b}|2\} 
\end{align}
\begin{align}\label{5b|1} b\neq -2,-1 :  \{5;b|1\}\cong\{5;\frac{-b}{1+b}|1\}\cong\{5;b|2\}\cong\{5;\frac{-b}{1+b}|2\} 
\end{align}
\begin{align}\label{60b1} b\neq 1 :\{6_0;b|1\}\cong \{6_0.iii2;b|2\}, \end{align}
\begin{align}\label{70b1} b\neq 1 :
\{7_0;b|1\}\cong \{7_0.i;b|2\},\cong \{7_0.ii;b|2\}
\end{align}
%\begin{align}\label{31bis}\{3|1\}\cong \{3v|3\}\cong \{3x|3\}\cong \{4iv|3\}\cong \{5iii|%3\}\cong \{6_0iv|3\}\cong \{6_{a\pm}|3\}.\end{align}
Algebras $\{4;-2|1\},\ \{5;-2|1\},\ \{4;-2|2\},\ \{5;-2|2\}$ are not equivalent to any other.

Matrices $C$ satisfying the equivalence \cond s \eqref{tfnX} are given in \cite{pehla:pp as JL}. For reader's convenience we give below only those needed in the examples \ref{11spec} and \ref{21spec}.

Interesting point is that isomorphisms of the Leibniz algebras are richer than those of corresponfing \dd s. For example, differently from the Leibniz algebra $[1|1]$,  the Abelian \dd\ is not isomorphic to any other.

\subsubsection{$C$-\mtr ces for pluralities \eqref{1|1}}

\begin{equation}\label{11to321}
 \{1\,|\,1\}
\rightarrow \{3;-2\,|\,1\}\ : \ \ C= 
\left(
\begin{array}{cccccc}
 -2 & 0 & 0 & 0 & 0 & 0 \\
 0 & 0 & \frac{1}{2} & 0 & 1 & 0 \\
 0 & 0 & -\frac{1}{2} & 0 & 1 & 0 \\
 0 & 0 & 0 & -\frac{1}{2} & 0 & 0 \\
 0 & \frac{1}{2} & 0 & 0 & 0 & 1 \\
 0 & \frac{1}{2} & 0 & 0 & 0 & -1 \\
\end{array}
\right),
\end{equation}
\begin{equation}\label{11to511} 
\{1\,|\,1\}
\rightarrow\{5;-1\,|\,1\}\ : \ \ C= 
\left(
\begin{array}{cccccc}
 -1 & 0 & 0 & 0 & 0 & 0 \\
 0 & 0 & 0 & 0 & 1 & 0 \\
 0 & 0 & 0 & 0 & 0 & 1 \\
 0 & 0 & 0 & -1 & 0 & 0 \\
 0 & 1 & 0 & 0 & 0 & 0 \\
 0 & 0 & 1 & 0 & 0 & 0 \\
\end{array}
\right),
\end{equation}
\begin{equation}\label{11to12} \{1\,|\,1\}
\rightarrow \{1\,|\,2\}
 \ : \ \ C= 
\left(
\begin{array}{cccccc}
 1 & 0 & 0 & 0 & 0 & 0 \\
 0 & 1 & 0 & 0 & 0 & 0 \\
 0 & 0 & 1 & 0 & 0 & 0 \\
 0 & 0 & 0 & 1 & 0 & 0 \\
 0 & 0 & -1 & 0 & 1 & 0 \\
 0 & 1 & 0 & 0 & 0 & 1 \\
\end{array}
\right),
\end{equation}
\begin{equation}\label{11to512} \{1\,|\,1\}
\rightarrow \{5;-1\,|\,2\}\ : \ \ C= 
\left(
\begin{array}{cccccc}
 -1 & 0 & 0 & 0 & 0 & 0 \\
 0 & 0 & 0 & 0 & 1 & 0 \\
 0 & 0 & 0 & 0 & 0 & -1 \\
 0 & 0 & 0 & -1 & 0 & 0 \\
 0 & 1 & 0 & 0 & 0 & 1 \\
 0 & 0 & -1 & 0 & 1 & 0 \\
\end{array}
\right).
\end{equation}

\subsubsection{$C$-\mtr ces for pluralities \eqref{2|1}}

\begin{equation}\label{21to411} \{2\,|\,1\}
\rightarrow \{4;-1\,|\,1\}\ : \ \ C= 
\left(
\begin{array}{cccccc}
 0 & 0 & -1 & 0 & 0 & 0 \\
 0 & 0 & 0 & 1 & 0 & 0 \\
 0 & 0 & 0 & 0 & -1 & 0 \\
 0 & 0 & 0 & 0 & 0 & -1 \\
 1 & 0 & 0 & 0 & 0 & 0 \\
 0 & -1 & 0 & 0 & 0 & 0 \\
\end{array}
\right),
\end{equation}
\begin{equation}\label{21to2i1} \{2\,|\,1\}
\rightarrow \{2.i\,|\,1\}\ : \ \ C= 
\left(
\begin{array}{cccccc}
 0 & 0 & 1 & 0 & 0 & 0 \\
 1 & 0 & 0 & 0 & 0 & 0 \\
 0 & -1 & 0 & 0 & 0 & 0 \\
 0 & 0 & 0 & 0 & 0 & 1 \\
 0 & 1 & 0 & 1 & 0 & 0 \\
 1 & 0 & 0 & 0 & -1 & 0 \\
\end{array}
\right),
\end{equation}
\begin{equation}\label{21to2ii2} \{2\,|\,1\}
\rightarrow \{2.ii\,|\,1\}\ : \ \ C= 
\left(
\begin{array}{cccccc}
 0 & 0 & 1 & 0 & 0 & 0 \\
 1 & 0 & 0 & 0 & 0 & 0 \\
 0 & 1 & 0 & 0 & 0 & 0 \\
 0 & 0 & 0 & 0 & 0 & 1 \\
 0 & -1 & 0 & 1 & 0 & 0 \\
 1 & 0 & 0 & 0 & 1 & 0 \\
\end{array}
\right),
\end{equation}

\begin{equation}\label{21to322} \{2\,|\,1\}
\rightarrow \{3:-2\,|\,2\}\ : \ \ C= 
\left(
\begin{array}{cccccc}
 0 & 0 & -2 & 0 & 0 & 0 \\
 \frac{1}{2} & 0 & 0 & 0 & -2 & 0 \\
 -\frac{1}{2} & 0 & 0 & 0 & -2 & 0 \\
 0 & 0 & 0 & 0 & 0 & -\frac{1}{2} \\
 0 & -\frac{1}{4} & 0 & 1 & 0 & 0 \\
 0 & -\frac{1}{4} & 0 & -1 & 0 & 0 \\
\end{array}
\right),
\end{equation}
\begin{equation}\label{21to412} \{2\,|\,1\}
\rightarrow \{4;-1\,|\,2\}\ : \ \ C= 
\left(
\begin{array}{cccccc}
 0 & 0 & -1 & 0 & 0 & 0 \\
 0 & 0 & 0 & 1 & 0 & 0 \\
 0 & 0 & 0 & 0 & -1 & 0 \\
 0 & 0 & 0 & 0 & 0 & -1 \\
 1 & 0 & 0 & 0 & 1 & 0 \\
 0 & -1 & 0 & 1 & 0 & 0 \\
\end{array}
\right),
\end{equation}
\begin{equation}\label{21to4iii2} \{2\,|\,1\}
\rightarrow \{4iii;1\,|\,2\}\ : \ \ C= 
\left(
\begin{array}{cccccc}
 0 & 0 & 1 & 0 & 0 & 0 \\
 0 & 0 & 0 & 1 & 0 & 0 \\
 0 & 0 & 0 & 0 & -1 & 0 \\
 0 & 0 & 0 & 0 & 0 & 1 \\
 1 & 0 & 0 & 0 & -1 & 0 \\
 0 & -1 & 0 & -1 & 0 & 0 \\
\end{array}
\right).
\end{equation}

\end{document}